\documentclass[12pt]{article}
\usepackage{amssymb,amsmath,esint,titlesec}
\titleformat*{\section}{\large\bf}
\titleformat*{\subsection}{\normalsize\bf}

\begin{document}


\begin{titlepage}

\setlength{\baselineskip}{24pt}

                               \vspace*{0mm}

                             \begin{center}

{\large\bf  The Legendre Transform in Non-additive Thermodynamics and Complexity}

                            \vspace*{3.5cm}

              \normalsize\sf    NIKOLAOS  \  KALOGEROPOULOS$^\dagger$\\

                            \vspace{0.2cm}
                            
 \normalsize\sf Carnegie Mellon University in Qatar\\
                         Education City, PO Box 24866\\
                            Doha, Qatar\\

                            \end{center}

                            \vspace{3.5cm}

                     \centerline{\normalsize\bf Abstract}
                     
                           \vspace{3mm}
                     
\normalsize\rm\setlength{\baselineskip}{18pt} 

We present an argument which purports to  show that the use of the standard Legendre  transform  
in non-additive  Statistical Mechanics is not appropriate. 
For concreteness, we use as paradigm,
the case of systems which are conjecturally described by the (non-additive) Tsallis entropy.
We point out the form of the modified Legendre transform that should be used, instead,
in the non-additive thermodynamics induced by the Tsallis entropy.  
We comment on more general implications of this proposal  for the thermodynamics of ``complex systems".

                           \vfill

                                                                         
                             \vfill

\noindent\rule{10cm}{0.2mm}\\  
   \noindent $^\dagger$ \small\rm Electronic address: \ \  \  {\normalsize\sf   nikos.physikos@gmail.com}\\

\end{titlepage}
 

                                                                                \newpage                 

\rm\normalsize
\setlength{\baselineskip}{18pt}

                                              \section{Introduction.}

There is little doubt, that the Legendre(-Fenchel) transform is one of the  best known and most useful mathematical devices encountered in any parts of Physics. 
The transform  commands a prominent role in Classical Mechanics, in Thermodynamics, in Statistical Physics and in Quantum Field Theory, and their applications, to just 
name a few such occurrences. Despite such a central role, there is still a certain amount of mystery, in the eyes of beginning students and even of some practitioners, 
about the meaning, interpretation and use of the Legendre transform (see \cite{ZRM} for a  pedagogical exposition). As a result, the role and use of the 
Legendre transform in such areas remains largely unexamined, unquestioned and unchallenged. It is  willingly accepted by the persons using it as a very 
successful device; it has certainly been that for the last century and a half during which it has been extensively employed. \\

A re-consideration and re-evaluation of the foundations of Statistical Mechanics has been taking place for about a quarter-century. 
One part of such a re-consideration has been driven by the maturity of the subject, its widespread applicability and its current challenges. 
Part of such challenges is its domain of applicability to newly developed sub-fields of Physics and far beyond, many of which come under 
the general title of ``Complex Systems". Parallel to this development has been the 
introduction and development in Physics of entropic functionals which are different from the classical and immensely successful Boltzmann/Gibbs/Shannon (BGS) entropy.
The most prominent power-law functional, and certainly the simplest one in functional form (so far as the author knows), has been the entropic functional
introduced by \cite{HC} and further explored by \cite{Vaj, Dar, LN, CR, RC}. It was independently re-discovered, brought to the attention of the Physics community and 
further developed by C. Tsallis \cite{Ts}, his collaborators and many others. For this reason, it is  widely called ``Tsallis entropy" \cite{Ts, Ts-book} by its supporters 
and  its critics alike, in the Physics community at least.\\

Even though some initial indications exist for the applicability of the  Tsallis entropy to physical systems (we make no comments about its potential applicability outside 
the parts of Physics that we are familiar with), its full range of applicability, if any, remains a subject of continuing discussion. The thermodynamic formalism
related to the Tsallis entropy, mirroring that of the Boltzmann/Gibbs/Shannon entropy  was already developed mostly within the first fifteen  years  after  
the introduction to the Physics community of the Tsallis entropy \cite{Ts-book}.  
This formalism uses in the same crucial way the classical Legendre transform as one uses it in  the case of systems described by the BGS entropy \cite{Ts-book}.
 The use of the Legendre transform in the Tsallis entropy, or more generally in the  so-called 
``non-extensive" (a precise term should have been ``non-additive") Statistical Mechanics, is taken for granted, as in the BGS case.\\

In this work we present an argument that states that we should not use the ordinary Legendre transform in a formalism involving 
non-additive entropies, such as  the Tsallis entropy. Therefore the already existing thermodynamic formalism based on the Tsallis and the other non-additive 
entropies that have sprung up during the last three decades has to be worked out again as the currently employed one is inappropriate. \\
 
An initial indication about this statement, may be seen as purely subjective or aesthetic: as one can easily see \cite{Ts-book} many/most of the 
thermodynamic statements involving the Tsallis entropy are quite complicated, and arguably also quite ``inelegant". 
This can be attributed to the fact that the systems that the Tsallis entropy purports to describe are more complicated than the more familiar ones 
described by the BGS entropy.  To refute this statement, one could also point out that  some of the most successful models in Physics could not 
be possibly be considered ``elegant", but are still enormously successful: the Standard Model in Particle Physics, the $\Lambda$-CDM model of Standard Cosmology
with inflation, many of the continuous and lattice models used in Statistical Physics etc.\\     

Irrespective of aesthetic reasons however, our proposal contained this work, is based on solid mathematical facts: we claim that if we choose, for whatever reason, 
the Tsallis (or any other non-additive) entropy to describe the collective behavior of a physical system with many degrees of freedom, then we have to use a 
transform which is different from, in particular a modification of, the familiar Legendre transform.  We are forced to  take this 
path if we want to employ a mathematical device that is faithful to and preserves the essential properties of the Legendre transform used in the BGS entropy-based 
thermodynamics. \\

The drawback of this approach is that we lose the universal applicability of the Legendre transform, and we replace it with transformation(s) that are adapted to the 
particular form of the entropic functionals that are used in each particular case. It is certainly true that several such functionals may use the same modification of the 
Legendre transform, as can be seen in this work. Still, the universality and uniqueness of the Legendre transform in thermodynamics and Statistical Mechanics, 
at least, will be lost. We consider this to be a positive development, even though it may appear to decrease substantially our ability to make universal statements.   
Ultimately one should not forget that these are mathematical devices, and only the comparison of predictions made using them with actual experimental data 
can validate their use in physical models.\\

The present work relies on the combination of results from two mathematically rigorous pillars. 
First, it relies on the results of the theory of optimal transport \cite{Vil-book} and Wasserstein spaces 
\cite{Lott, Lott2},  obtained by many authors, especially as such results pertain to a synthetic definition of Ricci curvature for metric measure spaces 
\cite{Sturm1, Sturm2, LV, Vil-book, CM}. Second, it relies on results in Convex Geometry/Analysis in  Euclidean spaces as well as in Functional Analysis  
\cite{AKM, KlartMil, AM1, BorS, AM2, AM3, AM4, AM5} that were obtained during the first decade of the  present century. We adopt, in particular, 
the form of the generalized Legendre transform, for the Tsallis entropy case at least, which was proposed  in \cite{AKM} and whose uniqueness 
was established \cite{AM3}. \\

We commented earlier on the relation between the Tsallis entropy and optimal transportation in \cite{NK1} where we also provided 
an isoperimetric interpretation of the non-extensive parameter. The present work is in the same geometric spirit and expresses its arguments 
from the same perspective, although oriented in a different direction, as \cite{NK1}.\\   
 
In Section 2, we highlight the relation between the Tsallis entropy and convexity. In Section 3, we discuss the Legendre transform, its generalization/modification  
that we adopt for the Tsallis entropy-based non-additive Statistical Mechanics, its uniqueness, and its suitability for our purposes. 
We conclude in Section 4 where we also  speculate about the potential implications of the present proposal. \\
  

\section{On entropies and convexity.}

\subsection{Boltzmann/Gibbs/Shannon and Tsallis entropies.} 
As is well-known, the classical Boltzmann/Gibbs/Shannon (BGS) entropy functional is given, for a system modelled on a sample space $\Omega$ endowed a volume 
element  \ $dvol_\Omega$ \ as
\begin{equation}
      \mathcal{S}_{BGS} [\rho] \ = \ - k_B \ \int_\Omega \rho(x) \log \rho(x) \ dvol_\Omega 
\end{equation} 
Usually \ $\Omega$ \ is chosen to be a Riemannian or Finslerian manifold endowed with an appropriate metric  \cite{Klauder, Ohta} and \ $dvol_\Omega$ \ 
represents a judiciously chosen (unique, in the Riemannian case) volume element associated with such a metric. The Tsallis entropy is defined as
\begin{equation}
     \mathcal{S}_q [\rho] \ = \ - k_B  \cdot \frac{1}{1-q} \left\{ 1 - \int_\Omega [\rho(x)]^q \ dvol_\Omega \right\} 
\end{equation}   
where \ $q\in\mathbb{R}$ \ (or even \ $q\in\mathbb{C}$) is called ``non-extensive" or ``entropic" parameter. The Boltzmann constant will be set \  $k_B=1$, \  
for simplicity, in the rest of this work. In the above \ $\rho$ \ indicates the Radon-Nikod\'{y}m derivative of the statistical measure (assumed to be Borel-regular) \
$d\mu$ \ of \ $\Omega$ \ with respect to the volume element \ $dvol_\Omega$ \ of \ $\Omega$. \ Clearly 
\begin{equation}
        \lim_{q\rightarrow 1} \mathcal{S}_q [\rho] \ = \ \mathcal{S}_{BGS} [\rho]  
\end{equation}


\subsection{($\lambda$-)convexity.} 

One of the most fundamental properties of \ $\mathcal{S}_{BGS}$ \ is its convexity (more precisely: its concavity). 
It is probably no exaggeration to state that most properties of $\mathcal{S}_{BGS}$  that  contribute to its central role and 
experimental success are related to  its convexity/concavity properties. Not too surprisingly, relying on this property of \ 
$\mathcal{S}_{BGS}$ \ forms also a core of a set of proposals intended to extend/extrapolate the validity of Statistical Mechanics, 
 in the case of systems with long-range interactions or even few degrees of freedom, see for instance \cite{Kastner}.\\
      
We recall that a subset \ $A \subset \mathbb{R}^n$ \ is convex, if for any two \ $x,y\in A$, \ the line segment 
\begin{equation}
          tx + (1-t)y \in A, \ \ \ \ t\in [0,1]    
 \end{equation}
lies in \ $A$. \  Generalizing from sets to real functions on \ $\mathbb{R}^n$, \ we recall that a function \
 $f: \mathbb{R}^n \rightarrow \mathbb{R}\cup \{ +\infty\} $  \ is called convex if 
\begin{equation}
     f(tx+ (1-t)y) \ \leq \ tf(x) + (1-t) f(y),   \hspace{10mm}  \forall \ x,y  \in \mathbb{R}^n, \ \ \ \forall \ t\in [0,1]
\end{equation}   
This property is pictorially interpreted as showing that the graph of the function \ $f$ \  lies below the straight line joining 
any two of the graph's points. Expressed differently, it means that the epigraph  (set) of \ $f$ \ is a convex set.   
As is well-known from elementary Calculus, a criterion for the convexity of such \ $f$ \ is that its Hessian be non-negative definite, 
which is expressed in coordinates \  $(z^1, z^2, \ldots z^n )$ \ of \ $\mathbb{R}^n$, \ with \ $\partial_i  \equiv \partial / \partial z^i, \ \ i=1, \ldots, n$, \ as 
\begin{equation} 
         (\partial_i \partial_j f) \ \geq \ 0 
\end{equation}
A third, probably less known but equivalent, characterization of convexity on the interval $[0, 1]$ is given in terms of the one-dimensional Green's function 
``propagator", which is the inverse of the Laplace operator with Dirichlet boundary conditions on the unit interval. This kernel $G(s,t)$ is explicitly 
\begin{equation}
        G(s,t) \ = \ \left\{
                \begin{array}{ll} 
                        t(1-s),  & \mathrm{if} \ \ s\leq t\\
                        s(1-t),  & \mathrm{if} \ \ s\geq t
                 \end{array}
                       \right.
\end{equation}
Then, for all sufficiently smooth functions \  $f: [0,1] \rightarrow \mathbb{R}$ \ such that their second derivative has a lower bound, one gets 
\begin{equation}
       f(t) \  \leq \  t f(0) + (1-t) f(1) - \int_0^1 \frac{d^2f(s)}{ds^2} \ G(s,t) ds 
\end{equation}
A function \ $f: \mathbb{R}^n \rightarrow \mathbb{R}$ \ is called concave is \ $-f$ \  is convex.  A non-negative function \
$F: \mathbb{R}^n \rightarrow \mathbb{R}$ \ is called logarithmically concave (``log-concave"), if \ $- \log F$ \  is convex. 
Hence a log-concave function \  $f: \mathbb{R}^n \rightarrow [0, +\infty)$  \  has the form
\begin{equation}
    e^{-\phi}, \ \ \  \phi : \  \mathrm{Convex}
\end{equation}
This definition shows that there is an injective correspondence between convex and log-concave functions.  
Log-convexity will be further discussed  in the next  Section.  \\

A generalization of convexity arises, if ones considers a lower bound on the Hessian (6) which is different from zero: let \
$f: \mathbb{R}^n \rightarrow \mathbb{R}$ \ be such that 
\begin{equation}
      (\partial_i \partial_j f) \ \geq \ \lambda g_{ij},   \ \  \ \ \ \ \lambda\in\mathbb{R}
\end{equation}
where \ $\mathbf{g}$ \ is the Riemannian metric tensor on \ $\mathbb{R}^n$ \ with coordinate components 
\begin{equation}
        \mathbf{g}(\partial_i, \partial_j) \ = \ g_{ij}
\end{equation}
Normally \ $g_{ij} = \delta_{ij}$ \ (the latter symbol being the Kronecker delta) 
for the Euclidean metric, but being more general than that does not affect any statement in what follows.  As is well-known, condition (9)
implies that 
\begin{equation}  
     f(tx+(1-t)y) \ \leq \ t f(x) + (1-t) f(y) - \frac{\lambda}{2} \ t(1-t) \ d^2(x,y), \ \ \ \  \  \   \forall \ x,y \in \mathbb{R}^n  
\end{equation} 
where \ $d(x,y)$ \ indicates the distance between \ $x$ \ and \ $y$ \ with the metric \ $\mathbf{g},$ \ and \ $t\in[0,1]$. 
One can see directly that $\lambda$-convexity of a function $f(x)$ in $\mathbb{R}^n$ amounts to the convexity of the function \ 
$f(x) - \frac{1}{2} d^2(0,x)$. \ Moreover $\lambda$-convexity is 
immediately generalizable to Riemannian manifolds as long as (11) is valid for at least one minimizing, constant-speed geodesic 
joining \ $x$ \ and \ $y$. \ The definition (10) can be further generalized by replacing \ $\lambda  g_{ij}$ \ by a quadratic form  \ 
$\Lambda$ \ on \ $T \mathbb{R}^n$ \ \cite{Vil-book}. \\   


 \subsection{Convexity of entropies.}
 
 To address the convexity properties of \ $\mathcal{S}_{BGS}$ \ and \ $\mathcal{S}_q$, \ one informally pretends
 that these functionals are actually  functions, observes that they are twice differentiable,  differentiates them twice  and uses the one-dimensional version of criterion 
 (6) for convexity. More formally one can functionally differentiate such expressions, but then appropriate statements have to be made about the properties of the 
 (infinite dimensional) space of  ``acceptable" functions $\rho$  entering (1) and (2). One way to sidestep such functional analytic subtleties that
 have to be addressed at some point, is to consider instead 
 the real functions \ $S_1 : \mathbb{R}_+ \rightarrow \mathbb{R}$ \ and \ $S_q : \mathbb{R}_+ \rightarrow \mathbb{R}$, \ defined by 
 \begin{equation}      
              S_1 (r) \ = \   r \log r, \ \ \ r > 0
 \end{equation}
and 
\begin{equation}
            S_q (r) \ = \   \frac{1}{1-q} (r - r^q), \ \ \ r>0, \ \ q\in\mathbb{R} 
\end{equation}
determine their convexity properties, and then infer the pertinent convexity/concavity properties (for our purposes) of the functionals (1), (2). This  is 
the way adopted in most  treatments of these issues in the Physics community. The indices $1$ and \ $q$ \ in (13), (14) are chosen to be compatible with the definitions 
(1), (2). Notice that there is an overall sign difference between (1), (2) and (13), (14) chosen so that the forthcoming results are expressed in the usual
notation  of the optimal transport community \cite{Vil-book}. This overall sign can be accounted for by substituting ``concave" for ``convex" for results 
stated in this Section, in the sequel. \\

Given the approach of the above paragraph, one can easily check that \ $-S_1(r)$ \ is a non-negative, concave function. 
Similarly, one can see that \ $-S_q(r)$ \  is also non-negative  for all \ $q\in\mathbb{R},$ \ and that  
\begin{equation}
              - S_q(r) \  \mathrm{is:} \   \left\{
         \begin{array}{ll}
                \mathrm{Concave,} & \mathrm{for} \ \ q>0\\
                                   &  \\
                \mathrm{Convex,}   & \mathrm{for} \ \ q<0
         \end{array}
            \right. 
\end{equation} 
 The \ $q=0$ \ case is of no particular physical interest, as \ $S_q(r)$ \ is a constant. Moreover, we observe that 
 \ $S_q(0) = 0$ \ for \ $q>0$, \ which is the case in which most of the conjectured applications of the Tsallis entropy seem to arise \cite{Ts-book, Ts2016}.
 In addition,  when $q<1$, one can easily see that the Tsallis entropy satisfies 
 \begin{equation}
     \mathcal{S}_q (A \cup B) \ \geq \mathcal{S}_q(A) + \mathcal{S}_q(B)
 \end{equation}
 for independent systems \ $A, B$, \ namely systems whose probability distribution functions satisfy 
 \begin{equation}
             p(A \cup B) \ = \ p(A) p(B)
 \end{equation}  
 In the above, $A \cup B$ indicates the system resulting from the combination of $A$ and $B$.
 The super-additivity property (20) of \ $\mathcal{S}_q$ \ is consistent with our experience with the increase of entropy upon combination of the underlying systems 
 enncoutered for systems described by \ $\mathcal{S}_{BGS}$. \ Therefore we consider such super-additivity to be a desirable result that \ $\mathcal{S}_q$ \ should obey. For this reason we choose to work  only with systems described by \ $\mathcal{S}_q$ \ for which \  $q<1$.  \ To conclude, we are interested, in this work, in  \ $S_q(r)$ \ which is a convex function, such that \ $S_q(0)=0$ \ and confine ourselves to the entropic parameter range  \ $q\in[0,1]$. \\
  
     
\subsection{Displacement convexity classes of functions.} 

In order to use the same notation (but not terminology) 
as several works in the field of optimal transportation,  such as \cite{Vil-book, LV} (and references therein) on which this Subsection relies, 
one introduces the parameter \ $N\in [1, \infty]$ \ which is given in terms of the entropic parameter \  $q\in [0,1]$  \  by
\begin{equation}
     N = \frac{1}{1-q}
\end{equation}
For reasons that would take us too far afield to motivate and explain in detail \cite{Vil-book}, and would not particularly contribute toward the argument of the present work, 
\cite{McCann} introduced the idea of ``displacement convexity" classes of functions \ $\mathcal{DC}_N$ \ which are labelled by the the \ $N$ \  
introduced in  (18), as follows \cite{Vil-book}: consider the set of continuous, convex functions \ $U: \mathbb{R}_+ \rightarrow \mathbb{R}$ \  
which are sufficiently smooth (being at least twice continuously differentiable turns out to be sufficient) on \ $(0, \infty)$ \ such that \ $U(0) = 0$. \ 
Moreover, one defines the ``topological pressure"  (more accurately, from a physical viewpoint, it should be called: the free energy)
\begin{equation}
        p(r) \ = \ r \frac{dU}{dr} - U(r)
\end{equation}
and by a second iteration of this formula
\begin{equation}
  p_2(r) \ = \ r \frac{dp}{dr} - p(r)
\end{equation}
Moreover, assume that the functions \ $U(r), \ p(r), \ p_2(r)$ \ satisfy one of the three  equivalent conditions:
\begin{enumerate}
    \item[$-$]  \ \   $p_2 + \frac{p}{N} \ \geq \ 0 $
    \item[$-$]  \ \  $p(r)/r^\frac{N-1}{N}$ \ is a nondecreasing function of \ $r$
    \item[$-$] \ \  $u(\delta)$ is a convex function of $\delta$, \ where \ \ \  $u(\delta) \ = \ \left\{ 
                        \begin{array}{lll}
                            \delta^N \ U(\delta^{-N}), & (\delta > 0), & \mathrm{if} \ \ N<\infty \\
                                         &   \\
                            e^\delta \  U(e^{-\delta}),  & (\delta\in\mathbb{R}), & \mathrm{if} \ \ N = \infty    
                        \end{array}
                                 \right. $                                
\end{enumerate}
Then all such functions \ $U$ \ belong to / constitute the displacement convexity class \ $\mathcal{DC}_N$. \\

The most studied concrete example of an element of \ $\mathcal{DC}_N$ \  is 
\begin{equation}   
  U_N (r) \ = \ \left\{
                  \begin{array}{ll}
                        N (r - r^\frac{N-1}{N}), & \ \ \ \mathrm{if} \ \ \ 1 < N < \infty \\
                              &   \\
                        r\log r, & \ \ \  \mathrm{if} \ \ \  N=\infty 
                  \end{array}
               \right.
\end{equation}
which upon using (18) gives  (14), for $q\in (0,1]$, and (13) respectively. By using this identification, we can  avoid using the more elaborate functional formalism
to show how convexity, and the displacement convexity classes \ $\mathcal{DC}_N$ \ are  intimately related to  \ $\mathcal{S}_{BGS}$ \ and \ $\mathcal{S}_q$. \
As can be seen by an explicit calculation, (21) satisfies \  $p_2 + p/N = 0$, \ so it saturates the lower bound of the first of the three defining relations of \ $\mathcal{DC}_N$. \
The essence of (21) and its use in the case of non-additive entropies lies in its power-law  character: consider, for instance, the function 
\ $\tilde{U} = r^d$ \ with \ $d\in\mathbb{R}_+$. \ We see that if \ $d\geq 1$, \ then such \ $\tilde{U}(r)$ \ belongs to all \ $\mathcal{DC}_N$ \ classes. \
So, from the viewpoint of the displacement convexity classes, the cases of \ $q > 1$ \ for the Tsallis entropy \ $\mathcal{S}_q$ \ are probably 
``too simple" to be of substantial interest.  If, on the other hand, \ $0 < d < 1$ \ then \ $\tilde{U} \in \mathcal{DC}_N$ \ if and only if \ 
\begin{equation}
      d \  \geq  \  \frac{N-1}{N}
\end{equation}
 So the function $\tilde{U}(r)  = r^\frac{N-1}{N}$ giving rise to an ``un-normalized" power law (Tsallis) entropy, is in some sense, the ``minimal representative" 
 of \ $\mathcal{DC}_N$ \ \cite{Vil-book}.  One can also remark that given (18), we can express (22)  as
 \begin{equation}      
           d \geq q
 \end{equation}
 In the same spirit, one can see that \  $U_\infty (r)$ \ in (21) is the limit of \ $U_N(r)$ \ as \ $N\rightarrow \infty$, \ which expresses in different terms the 
 well-known limit (3). \\       

Some properties of the $\mathcal{DC}_N$ classes pertinent to our discussion are:   
\begin{itemize}
        \item[$\circ$] We have the sequence of inclusions: \ \ \ $\mathcal{DC}_\infty \subset \ldots \subset \mathcal{DC}_3 \subset 
                                                          \mathcal{DC}_2 \subset \mathcal{DC}_1$ \\
                             Therefore, the smallest displacement convexity class is \ $\mathcal{DC}_\infty$ \ and the largest is \ $\mathcal{DC}_1$. 
        \item[$\circ$] ``Gauge freedom" for finite $N$: if  $U(r) \in \mathcal{DC}_N$ then \ $c_1U(c_2 r)+c_3 \in \mathcal{DC}_N$ for 
                                                                 any  $c_1\geq 0, \ c_2>0, \ c_3 \in \mathbb{R}$. \  
                             This shows that there is a 3-parameter family of possible entropic functionals that have the same convexity properties 
                             as the Tsalls entropy, all of which are equivalent from the displacement convexity viewpoint.  The Tsallis entropy uses  
                             ad hoc, but well-established from prior experience normalizations \cite{Ts-book} to fix \ $c_1$ \ and \ $c_3$. \ Fixing \ $c_2$ \ 
                             is trickier though, as this ultimately reflects a choice of a probability distribution in the phase space of the system. For ergodic
                             systems described by $\mathcal{S}_{BGS}$ this may be straightforward, due to ergodicity. But without a clear idea about the 
                             properties of phase space evolution of the systems described by the Tsallis entropy, determining \ $c_2$ \ a priori 
                             appears to be difficult.   
                                                   
        \item[$\circ$] ``Gauge freedom" for $N=\infty$: \ if \ $U(r) \in \mathcal{DC}_\infty$ \ then either $U(r)$ is linear, or there are constants \  
                             $c_4>0, \ c_5\in\mathbb{R}$ \ such that for all \ $r > 0$ \ one has \ \ $U(r) \ \geq \ c_4 \ r \log r + c_5 \ r$                                          
\end{itemize}
Further properties of these classes of functions can be found in \cite{Vil-book}. The above show that elements of \ $\mathcal{DC}_N$ \ can grow 
as slowly as we want at infinity for \ $N< \infty$, \ but have to grow at least as \ $r\log r$ \ for \ $N=\infty$. \ From such an asymptotic viewpoint, 
the functional form in \ $\mathcal{DC}_\infty$ \ giving rise to \ $\mathcal{S}_{BGS}$ \ is optimal.   \\    

At this point one can return from the functions (21)  or (13), (14) to the functionals (1), (2) whose convexity properties the forner  were defined to explore. 
As mentioned above, the problem with (1), (2) as opposed to (13), (14) is that the former are defined on spaces of probability measures of \
$\mathbb{R}^n$, \ usually with additional conditions on their moments or marginals to guarantee convergence. If one carries carefully this analysis 
\cite{Vil-book}, it turns out that the conditions defining \ $\mathcal{DC}_N$ \ are equivalent to demanding the corresponding functionals (1), (2) 
to be $\Lambda-$ convex in the infinite dimensional space of probability measures of \ $\mathbb{R}^n$ \ having finite first and second moments. 
In such a case \ $\Lambda$ \ stands for an appropriately defined  quadratic form, as mentioned above, on this space of measures \cite{Vil-book}. 
Hence  $\lambda$-convexity (10) 
is crucial as it is intimately connected to the definition of the classes of \ $\mathcal{DC}_N$ \ and \ $\mathcal{DC}_\infty$. In the sequel we will 
return to a typical ``naive", but quite common approach  to the problem: we will deal with such $\lambda-$ convexity issues in the finite-dimensional 
context of  \ $\mathbb{R}^n$, \ and functions such as (13), (14) rather than the infinite dimensional space of probability measures on $\mathbb{R}^n$ 
and the corresponding functionals (1), (2).  \\   


\section{Convexity and generalized Legendre transforms.}


\subsection{On log-concave measures and functions.} 

The importance and very wide applicability of the  log-concave functions in Geometry stems from the fact that the characteristic functions of 
convex sets in \ $\mathbb{R}^n$ \ have this property.  This should be intuitively/visually  obvious in low (2 or 3) dimensional Euclidean spaces. 
It can proved/quantified in any dimension, a variety of ways, many of which stem from 
the Br\"{u}nn-Minkowski inequality for the volume  of convex sets in \ $\mathbb{R}^n$: \ consider two compact sets \  $A, B \subset \mathbb{R}^n$ \ 
(not necessarily convex, but  we will confine ourselves to convex sets only, in this work) and let \ $\oplus$ \ stand for the Minkowski sum operation:
\begin{equation}
  A \oplus B \ = \ \{ x + y \in \mathbb{R}^n, \ \ \  x\in A, \ \  y \in B \} 
\end{equation}
and let \ $tA$ \ indicate the \ $t$-homothety of \ $A$
\begin{equation}
     tA \ = \ \{tx, \ \  x\in A, \ \ t\in \mathbb{R} \}
\end{equation}
If \ $vol$ \ stands for the Euclidean volume (Lebesgue measure), the Br\"{u}nn-Minkowski inequality \cite{Schneider} states that 
\begin{equation}
   \{ vol (A \oplus B) \} ^\frac{1}{n} \ \geq \ \{ vol (A) \} ^\frac{1}{n} + \{ vol (B) \} ^\frac{1}{n}
\end{equation}
which can be re-expressed as 
\begin{equation}
   \{ vol (tA \oplus (1-t)B) \}^\frac{1}{n} \ \geq \ t \{ vol (A) \}^\frac{1}{n} + (1-t) \{ vol (B) \}^\frac{1}{n}, \ \ \ \ \ t\in [0,1] 
\end{equation}
or, in dimension-independent form
\begin{equation}
    vol (tA \oplus (1-t)B) \ \geq \ \{vol (A) \}^t \ \{ vol (B) \}^{1-t}, \ \ \ \ \ t\in [0,1]
\end{equation}
which demonstrate that  the volume \ $vol$ \  is log-concave. Due to its significance, the Br\"{u}nn-Minkowski inequality has been extensively studied 
and  generalized in a variety of directions; there is a huge body of literature on the subject \cite{Schneider} 
 the most pertinent aspects of which, for our purposes, is the generalization of this inequality and its functional generalization (immediately the Pr\'{e}kopa-Leindler,  
 and even more generally the Brascamp-Lieb inequalities) 
 to Riemannian manifolds with lower Ricci curvature bounds and to metric-measure spaces with properties which turn out to generalize the lower Ricci curvature 
 bounds of the Riemannian case  \cite{Vil-book, EMil1, EMil2, EMil3, CM}.\\
 
 One can generalize the above statements from the volume \ $vol$ \  to any measure \ $\mu$ \ on 
 \ $\mathbb{R}^n$: such a measure is called log-concave, if for any measurable \ $A, B \subset \mathbb{R}^n$ \ and for any \ $t\in (0,1)$, \ we have
 \begin{equation}     
        \mu (tA \oplus (1-t)B) \ \geq \  \{ \mu(A) \}^t \{ \mu (B) \}^{1-t}
 \end{equation}
 Log-concave measures are the uniform measures of convex bodies. In addition, the Br\"{u}nn  concavity principle states that any lower dimensional marginal 
 of a uniform measure on a convex body is a log-concave measure.  These marginals are the only source of log-concave measures, as they form a dense 
 subset in the class of log-concave measures. Incidentally, an important example of log-concave measures are the Gaussians.\\  
 
 Further generalizing from measures to functions, a function \ $f: \mathbb{R}^n \rightarrow [0, \infty)$ \ is called log-concave if \ $\log f$ \ is concave on the 
 support of \ $f$. \  It is known \cite{Bor}, that a measure on $\mathbb{R}^n$ whose support in not contained in any affine hyperplane is log-concave if and 
 only if it is absolutely  continuous with respect to the volume (Lebesgue measure) and its density is a log-concave function.
 It is well-known that log-concave measures have  sub-exponential tails. To  deal with probabilities that have tails which are heavier than exponential, 
 \cite{Bor} introduces the classes of $s-$concave measures and $\kappa$-concave functions for \ \  $s, \kappa \in \mathbb{R}\cup \{ \pm\infty \}$.\\


\subsection{On \  $s$-concave measures and \ $\kappa$-concave functions.} 

By analogy and extending the definitions of the log-concave case, \cite{Bor}
states the following. Let \ $s\in [-\infty, +\infty]$, \ and \ $A, B \subset \mathbb{R}^n$ \ compact sets of positive measure \ $\mu(a)\mu(B) >0$. \ 
Let also \ $t\in[0,1]$. \ Then, the measure \ $\mu$ \ is called $s$-concave if  
 \begin{equation} 
      \mu(tA\oplus (1-t)B) \ \geq \ \left\{ 
                      \begin{array}{ll}
                          \left[ t \mu^s (A) +(1-t) \mu^s (B) \right]^\frac{1}{s},  & s \in \mathbb{R} \backslash \{ 0 \}  \\
                                            &          \\
                           \mu^t(A) \ \mu^{1-t}(B),                             &  s = 0 \\ 
                                            &          \\
                           \min  \{ \mu(A), \mu(B) \},                         &  s = -\infty\\
                                            &            \\
                           \max \{ \mu(A), \mu(B) \} ,                        &  s = +\infty
                      \end{array}
                             \right.
 \end{equation}
 Notice that if a measure is \ $s_1$-concave, it is also\  $s_2$-concave for \ \ $s_2 < s_1$. \\
 
  An extension from measures to non-negative functions on \ $\mathbb{R}^n$, \ goes as follows. 
 Let \ $\kappa \in [-\infty, +\infty]$. \ A function \ $f:\mathbb{R}^n \rightarrow \mathbb{R}_+$  \ is called  \ $\kappa$-concave,  if for all 
 \ \ $x,y\in\mathbb{R}^n, \ \  t\in [0,1]$
 \begin{equation} 
      f(tx+(1-t)y) \ \geq \ \left\{ 
                      \begin{array}{ll}
                          \left[ t f^\kappa (x) +(1-t) f^\kappa (x) \right]^\frac{1}{\kappa},  & \kappa \in \mathbb{R} \backslash  \{ 0 \} \\
                                            &          \\
                           f^t(x) \ f^{1-t}(y),                             &  \kappa = 0 \\ 
                                            &          \\
                           \min  \{ f(x), f(y) \},                         &  \kappa = -\infty\\
                                            &            \\
                           \max \{ f(x), f(y) \} ,                        &  \kappa  = +\infty
                      \end{array}
                             \right.
 \end{equation}
 According to the above definition, the log-concave functions are 0-concave. 
 One notices that \ $\kappa_1$-concave functions are also \ $\kappa_2$-concave if \ $\kappa_1 > \kappa_2$. \  The 
 Br\"{u}nn-Minkowski inequality can also be interpreted  as stating that the Euclidean volume is a \ $(1/n)$-concave function. 
 A concrete example of a \ $(-1/(n-1))$-concave function frequently  encountered in Physics is the Cauchy(-Lorentz) probability distribution in \ 
 $\mathbb{R}^n$. \\
 
 In \cite{Bor} a theorem was established relating $s$-convex measures and their \ $\kappa$-concave densities, 
 mirroring the case of log-concave  measures and their densities stated in the previous Subsection.  It states the following: let \ $\mu$ \ be a 
 measure on \ $\mathbb{R}^n$. \ Let \ $E$ \ be the affine subspace spanned by the support of \ $\mu$ \ with \ $\mathrm{dim} \ E = d$. \  Then, 
 for every \ $s\in [-\infty, 1/d]$, \ $\mu$ \ is \ $s$-concave if and only if it has a locally integrable density \ $\rho \in L^1_{loc} (\mathbb{R}^n)$, \ 
 where \ $\rho$ \ is \ $\kappa$-concave with \ $\kappa \in [-1/d, +\infty]$ \ given by 
 \begin{equation}  
       \kappa \ = \ \frac{s}{1-sd}
 \end{equation}
 In case \ $s> 1/d$, \ then \ $\mu$ \ is \ $s$-concave if and only if \ $\mu$ \ is a Dirac delta measure.  \\


\subsection{Dualities and generalized Legendre transforms.}

The classical Legendre(-Fenchel) transform is a fundamental tool in numerous parts of Physics, see e.g.  \cite{Arnold, ZJ, ZRM}, and Mathematics, 
especially ones closely associated to convexity \cite{Rock}. To define it, consider the class of  lower semi-continuous, convex  functions \
$f: \mathbb{R}^n \rightarrow \mathbb{R} \cup \{ \pm \infty \}$, \ denoted by \ $Cvx(\mathbb{R}^n)$. \ For most physical applications, 
considering continuous rather than lower semi-continuous  functions is sufficient and this is what we will assume from now on for the rest of this work. 
 The Legendre transform \ $\mathcal{L}$ \  of such a function \ $f$ \ is defined by 
\begin{equation}
      (\mathcal{L} f) (x) \ = \ \sup_y \ (\langle x, y \rangle - f(y)) 
\end{equation}
where \ $\langle \cdot, \cdot \rangle$ \ stands for the Euclidean inner product in \ $\mathbb{R}^n.$ \ Some of the well-known properties of the Legendre 
transform are:
\begin{itemize}
      \item The Legendre transform of a function \ $f:\mathbb{R}^n \rightarrow \mathbb{R}$ \ is a convex function: \ $\mathcal{L}f \in Cvx(\mathbb{R}^n)$.
      \item The Legendre transform is a bijection on the space \ $Cvx(\mathbb{R}^n)$.
      \item The Legendre transform is idempotent on \ $Cvx(\mathbb{R}^n)$: \ $\mathcal{LL}f = f, \ \ f\in Cvx(\mathbb{R}^n).$ 
      \item The Legendre transform reverses the direction of inequalities: if $f_1  \leq f_2$, then  $\mathcal{L}f_1 \geq \mathcal{L}f_2.$  
      \item $\mathcal{L}f_1 + \mathcal{L}f_2 = \mathcal{L}(f_1 \Box f_2)$ where $\Box$ indicates the infimal convolution of the two functions,
                which is  defined by \ \  $(f_1\Box f_2)(x) = \inf \{f_1(y) + f_2(z): \ x = y+z \}$  
\end{itemize}
All these properties, with the possible exception of the last one, are very well-known and extensively used in Physics. Despite that, the definition of the 
Legendre transform appears somewhat mysterious. An obvious question that arose is to find the essential properties that it possesses that make it 
such a flexible, even unique,  construction. This question was addressed surprisingly recently: during the first decade of the current century. It was 
eventually settled in \cite{AM4}.  It was  found that  the Legendre transform is the only ``natural" transform that someone could associate with the 
duality of convex functions. More precisely \cite{AM4}:  Let \ $\mathcal{T}: Cvx(\mathbb{R}^n) \rightarrow Cvx(\mathbb{R}^n)$ \  be a transform satisfying
\begin{itemize}   
  \item $\mathcal{T}^2 f \ = \ f$
  \item If \ $f_1 \ \leq f_2$ \ then \ $\mathcal{T}f_1 \ \geq \mathcal{T}f_2$ 
\end{itemize}
Then there exists a constant \ $c_0\in\mathbb{R}$, \  $x_0\in\mathbb{R}^n$ \ and a symmetric transformation \ $A\in GL_n(\mathbb{R})$ \ such that 
\begin{equation} 
       (\mathcal{T}f)(x) \ = \  (\mathcal{L}f) (Ax + x_0) + \langle x, x_0 \rangle + c_0
\end{equation}
where \ $GL_n(\mathbb{R})$ \  indicates the general linear group of $n\times n$ matrices with real entries. This result essentially establishes that 
the Legendre transform is the unique involution that reverses direction in the set of convex functions on \ $\mathbb{R}^n$  \ up to a linear transformation. 
It is a very general and far-reaching result, since it relies on rather mild assumptions but still manages to show the unique place that the classical 
Legendre transform occupies among all possible self-maps of \ $Cvx(\mathbb{R}^n)$. \\

Thermodynamics as we know it is based on log-concave functions such as the BGS entropy. One can re-express the log-concave character of 
the underlying theory by establishing an explicit functional duality which reflects the polar duality of the underlying convex structures. 
Such a duality was  first proposed in \cite{AKM} for the class of upper semi-continuous log-concave functions on $\mathbb{R}^n$ and goes as follows:
let \ $f = e^{-\phi}$. \ Then its  (polar) dual is defined to be 
\begin{equation}
   f^\circ = e^{-\mathcal{L}\phi}
\end{equation}
  or in more explicit terms as 
\begin{equation} 
      f^\circ(x) \ = \ \inf_{y\in\mathbb{R}^n} \frac{e^{-\langle x, y \rangle}}{f(y)}
\end{equation}
We can immediately see that (36) is the same as the Legendre transform (33) for \ $\phi$ \ remembering that the Legendre transform interchanges
$\sup$ \ and \ $\inf$. \         
Due to above characterization and uniqueness result  of \ \cite{AM4}, \  the Legendre transform is uniquely suited to be the  
transform of choice operating on the thermodynamic functions in a BGS entropy based formalism. 
This can probably explain its suitability and immense success as a theoretical tool, at least at the formal level, in the thermodynamic formalism. \\

Since, as argued in the previous Section, the Tsallis entropy is intimately related to the displacement convexity classes $\mathcal{DC}_N, N\in [1, +\infty]$
which are equivalent to the class $s$-concave functions on $\mathbb{R}^n$, one may be tempted to ask: is there  a transform, that plays the role for
$s$-concave functions similar to the role played by the Legendre transform for convex functions? The result is again affirmative and was provided in \cite{AM3}. 
To state it,  let \ $Conc_s(\mathbb{R}^n)$ \ be the class of all upper semi-continuous, $s$-concave functions on \ $\mathbb{R}^n$. \ Before proceeding,
one should stop and notice the discrepancy between the notation of \cite{AM3} employed in this Section and that of the previous section. What 
\cite{AM3} calls $s$-concave functions should be more conventionally called $1/s$-concave functions as it assumes that $f^\frac{1}{s}$ is concave on its 
convex support. Having this notational difference in mind, we continue with the notation adopted by \cite{AM3} in this Section. In \cite{AM3}, the authors 
prove that if \ $\mathcal{T}: Conc_s(\mathbb{R}^n) \rightarrow Conc_s(\mathbb{R}^n)$ \  is a transform such that 
\begin{itemize}         
    \item $\mathcal{T}^2 f \ = \  f$    
    \item If \ $f_1 \ \leq \ f_2$ \ then \ $\mathcal{T}f_1 \ \geq \mathcal{T}f_2$
\end{itemize}
then there exists  a constant \ $c_0 \in\mathbb{R}$ \ and a symmetric  \ $A\in GL_n(\mathbb{R})$ \ such that 
\begin{equation} 
        (\mathcal{T}f)(x) \ = \ c_0 \inf_{\{y: \ f(Ay)>0\} } \frac{\left( 1-\frac{\langle x,y\rangle}{s} \right)^s_+}{f(Ay)}
\end{equation}
In this expression, which is the counterpart of (34), the notation \ $( \cdot )_+$ \ stands for $\max \{ \cdot, 0 \}$. We immediately see that 
the numerator of  (37) is  the multi-dimensional version of the $q$-exponential with \ $s=1/(1-q)$ \  
which plays a central role in the formalism of the Tsallis entropy \cite{Ts-book}. The latter is conventionally defined by 
\begin{equation}    
      e_q(x) \ = \ \left\{1+ (1-q)x\right\}^\frac{1}{1-q} _+,   \ \ \ q\neq 1
\end{equation}
The result of the theorem (37)  guarantees that there is essentially a unique ``natural" Legendre transform for the class of $s$-concave functions on 
$\mathbb{R}^n$. This is not (33), but instead it is given explicitly by 
\begin{equation}
   (\mathcal{L}_sf)(x) \ = \ \inf_{ \{y: \ f(y)>0\} } \frac{\left(1 - \frac{\langle x,y \rangle }{s}\right)^s_+}{f(y)} 
\end{equation}
Hence, we are led to the following fact: it is not natural form the viewpoint of the Legendre transform to use (33) for a Tsallis entropy-based thermodynamics. 
Instead, if one wants to remain faithful to the ``spirit", namely the features of the Legendre transform that practically uniquely characterize it, then 
for all calculations in a Tsallis entropy-based Thermodynamics, one should use (39) instead.  This is the  main point of the present work.\\     

\noindent We would like to present a few comments pertinent to (37) and (39). \\

\noindent{\bf I.} \ The first comment is about  the way that $s$-concave measures arise 
geometrically. This is a counterpart statement to the log-concave measures as marginals of convex bodies previously mentioned in Section 2. To get 
a geometric interpretation, assume that \ $s\in\mathbb{N}$. \ The Br\"{u}nn concavity principle applies in this case too, and states that a function on 
$\mathbb{R}^n$ is $s$-concave if and only if it is a marginal of a uniform measure of a convex body in \ $\mathbb{R}^{n+s}$. \\

\noindent{\bf II.} \ Second, we state \cite{AKM} the relation of the polar duality of a convex body with the modified Legendre transform (37). This 
relation relies on the Br\"{u}nn concavity principle. We start by observing that \
$\mathcal{L}_s f  \leq  f^\circ$. \ Now, we  recall that for a convex body \ $K$ \ in \ $\mathbb{R}^n$, \  namely a convex set with a non-empty interior,
its polar \ $K^\circ$ \ is defined as 
\begin{equation}
     K^\circ \ = \ \{ x\in\mathbb{R}^n : \ \sup_{y\in K} \langle x,y \rangle \ \leq \ 1 \}
\end{equation}
This is the set theoretic analogue of (36); (36) is the functional generalization of (40). There are many ways that someone can define polarity of convex bodies. 
However, it turned out that all these ways are equivalent, up to a choice of a symmetric linear transformation \cite{BorS}.    
Consider the set 
\begin{equation}
    K_s(f) \ = \ \{ (x,y) \in \mathbb{R}^n \times \mathbb{R}^s: |y| \leq f^\frac{1}{s} (\sqrt{s}x) \}
\end{equation}
with the provision that \ $\sqrt{s}x$ \ should belong to the closure of the support of \ $f$. \ Then, up to rescaling, the function \ $f$ \ is the marginal of the 
uniform measure of the convex body \ $K_s(f)$ \ on \ $\mathbb{R}^n$. \ It turns out \cite{AKM} that 
\begin{equation}  
    [K_s(f)]^\circ \ = \ K_s (\mathcal{L}_s(f))
\end{equation}
which is the sought after relation. \\

\noindent{\bf III.} \ We also recall a statement of Subsection 2.4 regarding the inclusion of displacement convexity classes: in the current 
terminology, one sees that as $s$ increases toward infinity, the class of $s$-convex functions increases. The union of all these 
$s$-convexity classes is a dense  subset in the space of all log-concave functions of $\mathbb{R}^n$. \ Expressed differently,, 
one can see that any $s$-concave function is log-concave. Moreover it is possible to approximate any log-concave function 
$f: \mathbb{R}^n \rightarrow [0, +\infty)$ by an $s$-concave function $f_s$ which is 
\begin{equation}   
       f_s(x) \ = \ \left( 1+ \frac{\log f(x)}{s} \right)^s_+ 
\end{equation}
The fact that $f$ is log-concave implies that $f_s$ is $s$-concave and that the limit $f_s \rightarrow f$ as $s\rightarrow \infty$ is uniform locally in 
$\mathbb{R}^n$. \ This precise statement encodes the well-known fact  that in the limit $q\rightarrow 1$ of the Tsallis entropy, one should recover
not just the entropy, but the whole structure of the BGS entropy-based thermodynamics. \\  

\noindent{\bf IV.} \  In our previous work
\cite{NK1}, we commented on the central role of the  Bakry-\'{E}mery-Ricci curvature on the phase space of a system of many degrees of freedom, 
whose collective behavior is described by the Tsallis entropy. That generalized N-Ricci curvature behaved as if it was defined on a manifold of Hausdorff 
dimension $N$ even though the dimension of the underlying topological manifold (phase space) was just $n<N$. The difference was present due to 
the non-Riemannian nature of the effective measure on the phase space which was used to define this generalized Ricci curvature. The conventional Ricci
curvature was recovered in the limit that the effective measure became the Riemanian volume. We ascribed this to the fact that systems described by the 
Tsallis entropy are ``complex", so they cannot be easily factorized to a subsystem under study and an environment. In other words, for such complex 
systems, any subsystem is correlated with its environment in non-trivial ways. This viewpoint is evident in the present treatment. We see that for 
systems described by the Tsallis entropy, the Legendre transform has an exponential (36) ``kernel". Hence many of the details, such as the spatial extent in phase 
space, the number of the effective degrees of freedom etc  of the surrounding system/thermostat are infinite. This is a manifestation that the underlying system 
has short-range  (exponentially decaying) correlations in phase space, so it cannot probe the entirely of its surroundings. By contrast, systems described by the 
Tsallis entropy have ``fat tails", so correlations decay in a power-law manner. As such they are always sensitive to the details of their environment since they can 
probe its entirety. This is evident in the form of the modified Legendre transform applicable to them: it has a power law form (39). Naturally as \ $s\rightarrow\infty$ \ 
someone recovers the conventional  Legendre transform, which mirrors the fact that when \ $q\rightarrow 1$, \ the Tsallis entropy ``reduces" to the BGS entropy.\\
               
\noindent{\bf V.} \  Our last comment is about the distinguished role of the Gaussians in convexity and a conjecture about whether there is a similar role 
reserved for the $q$-Gaussians in a Tsallis entropy-based thermodynamics. There are numerous ways to single out, characterize and  understand the 
central role that the Gaussian measure in \  $\mathbb{R}^n$ \ plays in Statistical Mechanics. From the viewpoint of the classical Legendre transform, the following 
result was initially derived for even log-concave functions \ $f$ \ by K. Ball; the current formulation and proof is the main point of \cite{AKM}, which we closely follow.
The statement itself is a functional version of the well-known Blaschke-Santal\'{o} inequality for convex bodies in \ $\mathbb{R}^n$. \ 
Let $f: \mathbb{R}^n \rightarrow [0, +\infty)$ be an integrable function, namely \ $f\in L^1(\mathbb{R}^n)$. \  
Define $\tilde{f}$ to be the translate of $f$ by a constant vector \ $x_0\in\mathbb{R}^n$, \ 
namely \ $\tilde{f}(x) = f(x-x_0)$. \ Then 
\begin{equation}     
        \left(  \int_{\mathbb{R}^n} \tilde{f} \ \right) \ \left( \int_{\mathbb{R}^n} \tilde{f}^\circ \right) \ \leq \ (2\pi )^n 
\end{equation}
For \ $f$ \ log-concave, one may choose \ $x_0$ \ to be the location of the ``barycenter", i.e. of the mean value of $f$, namely 
\begin{equation} 
        x_0 \ = \ \frac{\int_{\mathbb{R}^n} x f(x)}{\int_{\mathbb{R}^n} f(x)} 
\end{equation}
The minimum over \ $x_0$ \ of the left-hand-side (where such a minimum is called the Santal\'{o} point of $f$) equals the right-hand-side in (44) 
if and only if \ $f$ \ is Gaussian. This can be re-expressed as stating that 
\begin{equation}
      \left(\int_{\mathbb{R}^n} e^{-\tilde{f}} \right) \  \left(\int_{\mathbb{R}^n} e^{-\mathcal{L}(\tilde{f})} \right) \ \leq \ (2\pi )^n
\end{equation}   
One can generalize (46) to the case of $s$-concave functions, by using the  modified Legendre 
transform (39) instead of  (33).  Using  the notation of (38), (39) we  \emph{conjecture}  that for non-negative, \ $s$-concave functions \  
$f\in Conc_s (\mathbb{R}^n)$ \ we should have 
\begin{equation}  
          \left(  \int_{\mathbb{R}^n} e_q(-\tilde{f}) \right) \ \left( \int_{\mathbb{R}^n} e_q(-\mathcal{L}_q(\tilde{f})) \right) \ \leq \ C(q)
\end{equation}
where $C(q)>0$ is a constant, and that the equality should be attained by $q$-Gaussians. \\


\section{Discussion and conclusions}

In this work we developed an argument based on convexity that leads to the following proposal: if one wishes to use the Tsallis or other entropic
functionals to develop an appropriate thermodynamics for systems whose collective description is given by these functionals, \ \emph{one has no choice
but to abandon the ``usual" / well-known definition of the Legendre transform.} This modification is absolutely necessary, if such thermodynamics 
follows the spirit rather than the letter of what constitutes a Legendre transform. \emph{Therefore the existing Tsallis entropy-based thermodynamics 
\cite{Ts-book} has to be re-written, by using the stated in this work generalized/modified Legendre transform rather than the conventionally used one}.\\

Our argument, although non-rigorous itself,  relies on relatively recent and rigorous (which may even be well-known) mathematical results 
obtained in two different fields: on the one hand on the Optimal Transportation of Measures and the subsequent synthetic definition of Ricci curvature 
for metric measure spaces and on the other hand on results of Convex Geometry/Analysis in Euclidean spaces. Our contribution rests in putting 
such arguments together and looking at them under the viewpoint of a generalized/modified  thermodynamic formalism controlled/dictated by 
non-BGS entropies in general,  and by the Tsallis entropy, in particular.\\

\noindent Our proposal has the following potential implications:\\

\noindent {\bf 1.} \    The ``usual" / ``ordinary" Legendre transform loses its unique status in Statistical Mechanics and Thermodynamics. 
           In principle, for each different entropic functional employed one has to use the appropriate (and quite possibly uniquely defined)  Legendre transform.  
           On the other hand, different entropic functionals may use the same Legendre transform: as an example functionals descending from functions 
           belonging to the same displacement convexity class \ $\mathcal{DC}_N$ \ will have to use the same Legendre transform, according 
           to the arguments of this work. \\
            
\noindent {\bf 2.} \   What we propose in this work decreases substantially the universality and predictive power of Thermodynamics. 
            Thermodynamics has been   phenomenology 
           since its earliest days. Hence its robustness and power: one does not need to know all the details of the underlying system top predict its macroscopic 
            behavior, all that is needed are a few macroscopic parameters.   
           There is no need to know what exactly entropy is at the microscopic level to develop the formalism. It certainly has helped that only one entropy functional 
           (the BGS entropy) was considered until recently and the thought of using alternative functionals did not even arise (as it does not, even today) in 
           the thoughts of many practitioners. One should always remember that the statistical foundations of thermodynamics were developed having in 
           mind weakly interacting and ergodic systems in (quasi-) equilibrium
           such as ideal, or low density, gases. When one tries to use thermodynamics for more general systems, 
           such as systems with long-range interactions, with strong spatio-temporal  correlations, without  phase space ergodicity, systems out of equilibrium  etc, 
           then it is not obvious that the conventional thermodynamic concepts necessarily apply without modification or have any meaning \cite{Ts-book} at all. 
           One testimony toward this, is that  that there is still no consensus about what would constitute a reasonable formalism 
            of non-equilibrium thermodynamic,  despite several decades of effort in this direction. 
           The set of phenomena which are not necessarily described by the BGS entropy
           are probably too diverse/different from each other to be captured by the single thermodynamic formalism that has  been employed so far. Our argument 
           in this work clearly supports he idea that multiple thermodynamic formalisms should exist each one relying on its own entropy, associated Legendre 
           transform etc.    \\
                               
\noindent {\bf 3.} \     If the ``usual" / ``ordinary" Legendre transform loses its unique status in Statistical Mechanics and Thermodynamics and has to be modified depending 
           on the employed entropic functional, then what impact would that have in other parts of Physics, where it is extensively employed, such as Classical 
           Mechanics, for instance? After all, Classical Mechanics relies on Newton's equations aptly generalized in the Lagrangian or Hamiltonian formalisms.
           Newton's equation are of the second order and hence they give rise to quadratic kinetic terms in either one of these formalisms. 
           We attempted to connect this to the Euclidean nature of the spacetime metric in \cite{NK2}. How appropriate would it be to use the Legendre transform
           as is, for Lagrangians or Hamiltonians not having quadratic kinetic terms? Theoretical proposals for how to handle such ``exotic" systems have 
           abounded over the decades, and interest in them has mostly mathematical significance, as all fundamental physical theories (the Standard Model 
           of Particle Physics and General Relativity) involve quadratic kinetic terms. However models with such exotic kinetic terms can be used as effective 
           descriptions of systems when a more fundamental description is either not sought after or is just intractable.  The recent rise in popularity of quantization 
           approaches inequivalent to the conventional ones, encountered in Quantum Field Theories, involving by non-standard kinetic terms such as Chern-Simons 
           theories in Loop Gravity, and topological insulators, for instance,
           make answering the above questions of potentially immediate physical significance for fields outside conventional Thermodynamics.\\             
 
\noindent {\bf 4.} \   The Tsallis entropy conjecturally describes dynamical systems having weak mixing properties, fat tails in their probabilistic distributions etc.  
          All the definitions in dynamical systems theories that the author is aware of, explore in a very essential way the underlying hyperbolicity of the system 
          and their exponential instability  (``chaos"). Consequently all entropies employed have the BGS form. As as result, the whole thermodynamic formalism 
          employed in dynamical systems uses in a very essential way the usual definition of the Legendre transform. Employing the Tsallis entropy 
          in dynamical systems will force us  not only to use a different functional form for the entropies employed, but also a different Legendre transform, 
          thus requiring a radical re-consideration and extension of the overall formalism employed for dynamical systems.\\ 

\noindent {\bf 5.}  \   We are forced to re-consider the basis of dualities in the functional spaces that we use. An example is the Fourier transform. 
         There have been proposals of how 
         to generalize/adapt the Fourier transform to the case of the Tsallis entropy involving the $q$-deformed, instead of the ``usual", exponentials. The
         concrete  implementation has remained, due to some undesirable features, somewhat controversial \cite{UTS, Hil, JT, PR1, PR2, PR3}. 
         It may be worth revisiting this topic due, in no small part, to its centrality in performing concrete calculations,  
         adopting the viewpoint leading to the generalized/modified Legendre transforms employed in the present work.  \\

We conclude the present work, which is quite full of rather abstract statements, by reminding ourselves the following obvious facts, as part of an obvious but strong 
self-critique:  Physics is an experimental science, and the proposal of the present work has any physical merit if and only if it can give predictions that are compatible 
with experiment.   Hence the physical relevance and potential physical importance of the thesis put forth  in the present work, in the framework of Tsallis entropy,  
remains to be determined, after specific predictions are made based on concrete physical models employing the proposed approach. We intend to address such
points, and explore the extent  to which such predictions are analytically feasible/tractable, in future work. \\  




\end{document}